\documentstyle[11pt,newpasp,twoside,epsf]{article}
\def\etal{{\rm et al.\ }}

\def\lsim{\mathrel{\hbox{\rlap{\hbox{\lower4pt\hbox{$\sim$}}}\hbox{$<$}}}}
\def\gsim{\mathrel{\hbox{\rlap{\hbox{\lower4pt\hbox{$\sim$}}}\hbox{$>$}}}}

\def\edcomment#1{\iffalse\marginpar{\raggedright\sl#1\/}\else\relax\fi}
\reversemarginpar

\begin{document}

\title{The Inner Structure of Cold Dark Matter Halos}

\author{Julio F. Navarro\altaffilmark{1}}

\affil{Department of Physics and Astronomy, University of Victoria, Victoria,
BC, V8P 1A1, Canada}

\altaffiltext{1}{CIAR Scholar and Alfred P. Sloan Research Fellow}

\begin{abstract}{
I report on recent progress in our understanding of the structure of CDM halos,
and in particular of the inner mass profile of galaxy-sized systems. Numerical
simulations have consistently shown that the density profiles of CDM halos
steepen monotonically from the center outwards, with slopes shallower than
isothermal near the center and steeper than isothermal near the virial
radius. Ongoing debate centers on the precise radial dependence of the
logarithmic slope, as well as on whether it approaches a well defined asymptotic
central value. The latest high-resolution simulations suggest that the circular
velocity profile is well approximated by the model proposed by Navarro, Frenk \&
White (NFW). On the other hand, the radial dependence of the slope of the
density profile differs modestly, but significantly, from the model proposed by
NFW. As a result, NFW fits tend to underestimate the density at radii just
inside the scale radius.  Rather than implying a very steep ($\rho \propto r^{
-1.5}$) inner divergent slope, I argue that the data is actually best
represented by a model where the density profile becomes increasingly shallow
with radius, with little sign of approach to a well-defined asymptotic value.  A
model where the {\it phase-space density profile} is a power law accounts well
for these results and suggests that the innermost slope may be as shallow as
$\rho \propto r^{-0.75}$.  These conclusions are supported by a thorough
numerical convergence study that elucidates the effect of numerical parameters
such as the timestep, gravitational softening, and particle number, on the mass
profile of simulated dark matter halos.}
\end{abstract}

\section{Introduction}

Cosmological N-body simulations have led to impressive strides in our
understanding of structure formation in universes dominated by collisionless
dark matter. The impact of such simulations has been greatest in the highly
non-linear regime, where analytic calculations offer little guidance. Recently,
and as a result of the development of efficient algorithms and of the advent of
powerful, massively parallel computer arrays, it has been possible to apply
N-body studies to detailed investigations of structure on small scales. These
studies can now probe scales comparable to the luminous radii of individual
galaxies, thus enabling direct comparison between theory and observation in
regions where luminous dynamical tracers are abundant and easiest to
observe. Predicting the structure of dark matter halos on kpc and sub-kpc
scales, where it can be compared directly with observations of galactic
dynamics, is one of the premier goals of N-body experiments, and there has been
steady progress in this area over the past few years.

Building upon the early work of Frenk et al (1988), Quinn, Salmon \& Zurek
(1986), Dubinski \& Carlberg (1991) and Crone, Evrard \& Richstone (1993),
Navarro, Frenk \& White (1996, 1997, hereafter NFW) found that, independent of
mass and of the value of the cosmological parameters, the density profiles of
dark matter halos formed in various hierarchical clustering cosmogonies were
strikingly similar. This `universal' structure can be characterized by a
spherically-averaged density profile which differs substantially from the simple
power-laws, $\rho(r) \propto r^{-\beta}$, predicted by early theoretical studies
(Gunn \& Gott 1972, Fillmore \& Goldreich 1984, Hoffmann \& Shaham 1985, White
\& Zaritsky 1992). The profile steepens monotonically with radius, with
logarithmic slopes shallower than isothermal (i.e. $\beta < 2$) near the center,
but steeper than isothermal ($\beta>2$) in the outer regions.

NFW proposed a simple formula,
\begin{equation}
\label{eq:nfw}
{\rho(r) \over \rho_{\rm crit}} = {\delta_c \over (r/r_s)(1+r/r_s)^2},
\end{equation}
which describes the density profile of any halo with only two parameters, a
characteristic density contrast{\footnote{I use the term `density contrast' to
denote densities expressed in units of the critical density for closure,
$\rho_{\rm crit}=3H^2/8\pi G$. I express the present value of Hubble's constant
as $H(z=0)=H_0=100\, h$ km s$^{-1}$ Mpc$^{-1}$}}, $\delta_c$, and a scale
radius, $r_s$. Note that there is no well defined value for the {\it central}
density of the dark matter, which can in principle climb to arbitrarily large
values near the center. This is an important point, especially because there
have been a number of reports in the literature arguing that the shape of the
rotation curves of many disk galaxies rules out steeply divergent dark matter
density profiles (Flores \& Primack 1994, Moore 1994, McGaugh et al 1998, de
Blok et al 2001).

The results of NFW have been confirmed by a number of subsequent studies (see,
e.g., Cole \& Lacey 1996, Huss, Jain \& Steinmetz 1999, Jing \& Suto 2000),
although there is some disagreement about the innermost value of the logarithmic
slope. Moore et al.  (1998), Ghigna et al. (2000), and Fukushige \& Makino
(1997, 2001) have argued that density profiles diverge near the center with
logarithmic slopes steeper than the asymptotic value of $\beta=1$ in NFW's
formula.  Since steep inner slopes are apparently disfavored by rotation curve
data it is important to establish this result conclusively; if confirmed, it may
offer a way to falsify the CDM paradigm on small scales.

Unfortunately, observational constraints are strongest just where theoretical
predictions are least trustworthy. For example, the alleged disagreement between
observed rotation curves and cuspy dark halo models is most evident in
sub-$L_{\star}$ galaxies on scales of $\sim 1\, h^{-1}$ kpc or less. For typical
circular speeds of $\sim 100$ km s$^{-1}$, this corresponds to regions where the
density contrast exceeds $\sim 10^6$. Orbital times in these regions are of
order $10^{-3}$ of the age of the universe, implying that N-body codes must be
able to follow particles accurately for several thousand orbits. Few
cosmological codes have been tested in a systematic way under such
circumstances. Extreme care is thus needed to separate numerical artifacts from
the true predictions of the Cold Dark Matter model.

Furthermore, there are {\it no} simulations (with adequate resolution) of dark
matter halos with masses comparable to the dwarf galaxies where the disagreement
between theory and observation is most evident. So far, published work has
relied upon extrapolation of results obtained for massive galaxy halos and
galaxy clusters, hardly an ideal procedure for assessing the viability of the
ruling CDM paradigm on small scales. In order to validate or `rule out' the CDM
cosmogony one must be certain that model predictions on the relevant scales are
accurate, robust, and free of systematic numerical uncertainties.

\section{Numerical Convergence Criteria}

Motivated by this state of affairs, we have undertaken a large series of
numerical simulations designed to clarify the role of numerical parameters on
the structure of simulated cold dark matter halos. What regions of a simulated
dark matter halo in virial equilibrium can be considered reliably resolved? Our
tests explore the influence of the gravitational softening, the time-stepping
algorithm, the starting redshift, the accuracy of force computations, and the
number of particles on the spherically-averaged mass profile of a galaxy-sized
halo in the $\Lambda$CDM cosmogony. A thorough discussion of the tests is
presented in Power et al (2001); I summarize below a list of conditions that
must be satisfied in order to consider the mass profile reliably resolved at a
given radius:

\begin{itemize}

\item
The timestep must be substantially shorter than the orbital timescale:
$t_{\rm circ}(r)\gsim 15 \left(N_{\Delta t}\right)^{-5/6} t_{\rm
circ}(r_{200})$, where $N_{\Delta t}$ is the total number of timesteps, $t_{\rm
circ}(r)=2\pi r/V_c(r)$ is the circular orbit timescale, and $r_{200}$ is the
virial radius, where the mean density contrast is $200$. This criterion applies
provided that the gravitational softening spline scalelength, $\epsilon$, is
chosen to ensure that particle discreteness effects are negligible, $\epsilon
\gsim 4\, r_{200}/\sqrt{N_{200}}$, where $N_{200}$ is the number of particles
within the virial radius. Smaller softenings require {\it substantially smaller
timesteps} for convergence.

\item
Accelerations do not exceed a characteristic value imprinted by the
circular speed of the halo, $V_{200}$, and by the softening scale:
$a(r)=V_c^2(r)/r \lsim a_{\epsilon}=\chi_{\epsilon} V_{200}^2/\epsilon$, with
$\chi_{\epsilon}\approx 0.5$.

\item
Enough particles are enclosed so that the collisional relaxation
timescale, $t_{\rm relax}= (r/V_c(r)) \, N(r)/(8\ln{\Lambda_{\rm C}})$, is
comparable to the age of the universe, $t_0$. Empirically, we find that regions
where $t_{\rm relax} \gsim 0.3 \, t_0$ are adequately resolved. ($N(r)$ is the
enclosed number of particles and $\ln{\Lambda_{\rm C}}\approx \ln{N(r)}$ is the usual
Coulomb logarithm.)

\end{itemize}

\noindent
Convergence also requires that at the initial redshift, $z_i$, the linear rms
fluctuations on the smallest resolved mass scale (the particle mass $m_p$) is
$\sigma(m_p,z_i) \lsim 0.3$, and that force calculations are highly accurate.
Poor spatial, time, or force resolution leads generally to systems with
artificially low central density, but may also result in the formation of
artificially dense central cusps.  This feature must be monitored carefully in
hierarchical clustering simulations, since cusps formed in poorly-resolved
high-redshift progenitors may survive merging and compromise the inner structure
of present-day systems.

The most stringent requirement for convergence is that imposed on the particle
number by the collisional relaxation criterion, which implies that in order to
estimate accurate circular velocities at radii as small as $1\%$ of $r_{200}$,
where the density contrast may reach $\sim 10^6$, the region must enclose at
least $3,000$ particles (or more than a few times $10^6$ within the virial
radius). We can use these criteria to identify regions of simulated dark halos
where the mass profiles are unlikely to be affected by numerical artifacts.

\section{Phase-Space Density Profiles}

\begin{figure}
\plottwo{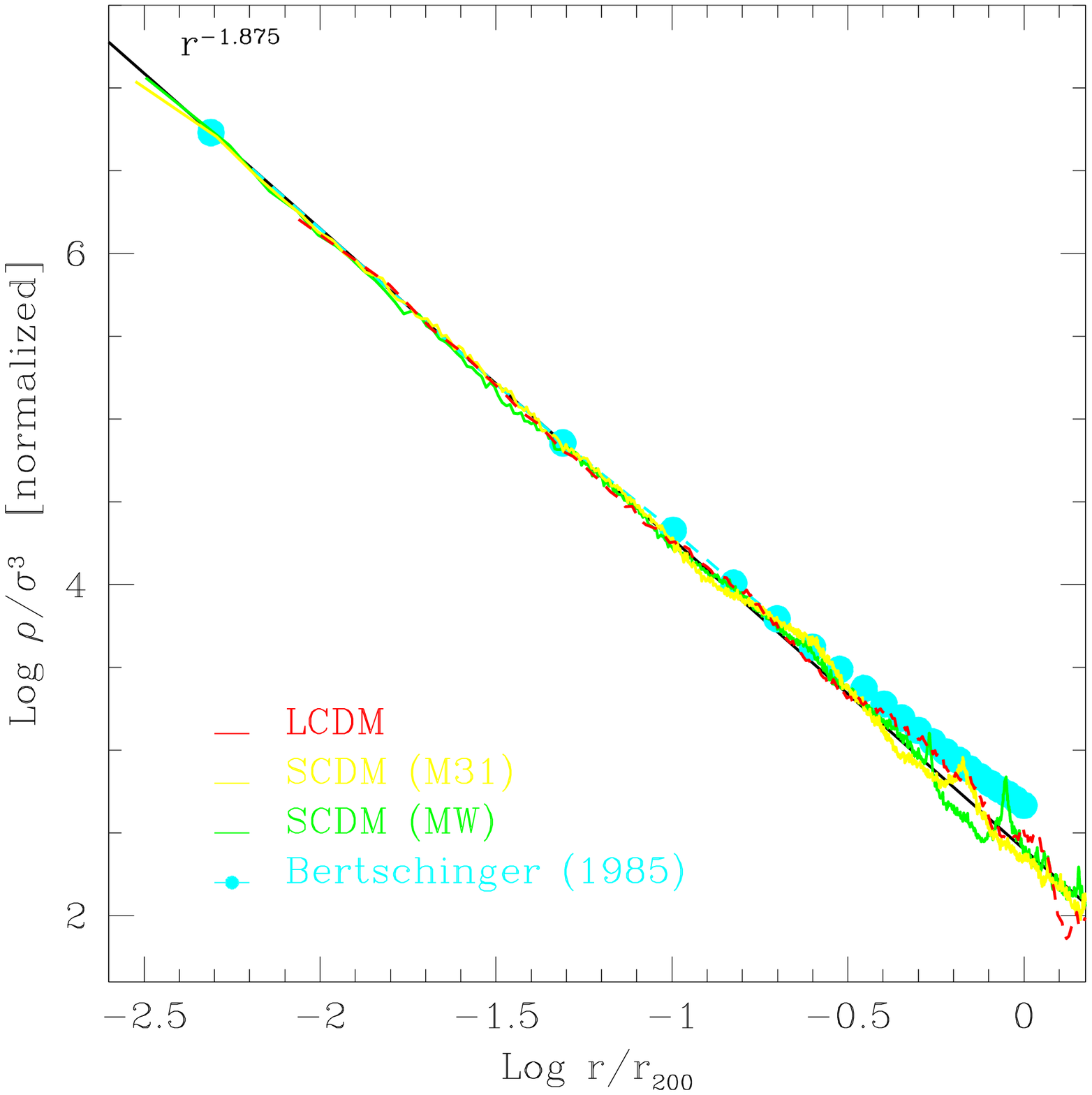}{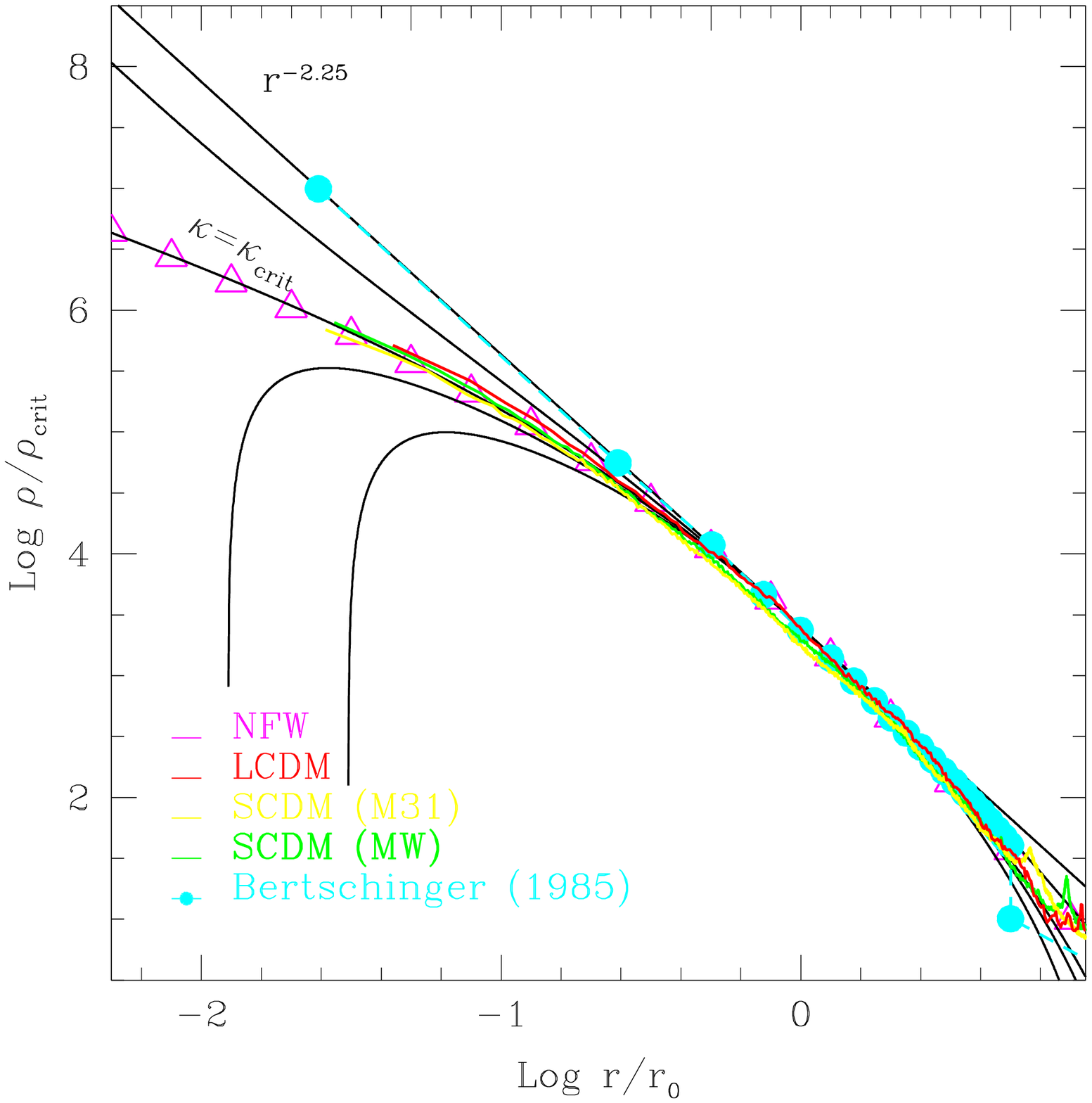}
\caption{
\small {\it Left (a):} The phase-space density profiles of three galaxy-sized CDM
halos. Solid lines correspond to the SCDM halos and dashed lines correspond to
the LCDM halo. Vertical normalizations are arbitrary and have been chosen so
that the curves coincide at about $0.01 \, r_{200}$. Radii are normalized to the
virial radius, $r_{200}$. The solid circles indicate the self-similar solution
obtained by Bertschinger (1985) for spherical infall of gas onto a point mass
perturber in a uniform Einstein-de Sitter universe. Radii for this solution have
been normalized by assuming that the shock radius in the solution equals
$r_{200}$. A power-law of slope $-1.875$ is shown for comparison (thin solid
line). {\it Right (b):} Density profiles of the CDM halos and the Bertschinger
solution, compared with analytic solutions to Jeans' equations derived using the
power-law entropy stratification constraint (solid continuous lines). Radii have
been scaled here to $r_0$, the radius where the slope of the density profile is
$-2.25$, and densities to the critical density. }
\end{figure}

As discussed in \S1, there is at present no broad consensus regarding how steep
the innermost slope of dark matter density profiles is, or even whether there is
a well-defined value for the asymptotic innermost slope. Unfortunately, the
results of the previous section make it clear that it will take extraordinary
computational effort to reach a robust resolution of the controversy. What is
required is a statistically significant sample of galaxy-sized halos simulated
with sub-kpc resolution; this is an extremely onerous computational task that
will stretch the capabilities of the most powerful massively parallel
computers. Steps in this direction are currently being taken, but it will take
some time until these efforts yield conclusive results.

From the theoretical point of view, a number of plausible arguments have been
advanced in order to try and explain the innermost behaviour of dark matter
density profiles from stellar dynamical principles. These efforts, however, tend
to give non-unique results and have so far been unable to explain the remarkable
similarity in the structure of dark matter halos of widely different mass formed
in a variety of cosmogonies (Evans \& Collett 1997, Syer \& White 1998, Nusser
\& Sheth 1999, Lokas \& Hoffman 2000).

We have recently explored an empirical alternative to analytic efforts addressed
at estimating the innermost slope of the density profile (Taylor \& Navarro
2001, hereafter TN). Our approach exploits the similarity between the {\it
phase-space density} profiles (defined here as the ratio of density to velocity
dispersion cubed, $\rho/\sigma^3$, measured in spherical shells) of CDM halos
and of the self-similar solution for spherical collapse in an expanding universe
found by Bertschinger (1985). This offers an attractive scenario for
understanding the shape of halo density profiles as well as a powerful tool for
estimating their slope near the center.

Figure 1a shows the spherically-averaged phase-space density profile of three
simulated CDM halos. Two of these halos have been identified in simulations of
the former ``standard'' CDM cosmogony (SCDM, $\Omega_m = 1, \Lambda = 0, h =
0.5, \sigma_8=0.7$) and have circular velocities of $\sim 180$ and $\sim 160$ km
s$^{-1}$, respectively. These two halos are part of the ``Local Group''
simulation reported by Moore \etal (1999). The third halo has a circular
velocity of $\sim 200$ km s$^{-1}$ and was run in the currently popular $\Lambda$CDM
cosmogony ($\Omega_m = 0.3, \Lambda = 0.7, h = 0.65, \sigma_8=0.9$). Solid
(dashed) lines are used for the SCDM ($\Lambda$CDM) halos.  For ease of comparison, we
have chosen to normalize $\rho/\sigma^3$ so that all three curves coincide at
$0.01 \ r_{200}$. All three systems have $\sim 10^6$ particles within $r_{200}$
and are amongst the highest resolution simulations available at present.  The
important point illustrated by Figure 1a is that, over more than two decades in
radius, the phase-space density profile is very well approximated by a power law
of slope $-1.875$ (solid straight line). This is quite remarkable, given that
both the density profiles (shown in Figure 1b) and the velocity dispersion
profiles of these halos deviate substantially from simple power laws.

Also remarkable is that this power-law coincides with the self-similar solution
derived by Bertschinger (1985) for secondary infall onto a spherical
perturbation in an unperturbed Einstein-de Sitter universe: Bertschinger's
solution is plotted with solid circles in Figure 1. The solution corresponds to
the self-similar equilibrium configuration of a $\gamma=5/3$ gas formed by
spherical accretion onto a point-mass perturber in an otherwise uniform
Einstein-de Sitter universe. The quantity shown by the solid circles in Figure
1a, $\rho^{5/2}/P^{3/2}$, is roughly equivalent to the phase-space density ($P$
is the local pressure). As discussed by Bertschinger, this solution is the most
appropriate to compare with our numerical results for CDM halos, given that the
velocity dispersion tensor in this case is only mildly anisotropic. Radii are
normalized assuming that $r_{200}$ equals the shock radius of the self-similar
solution, which corresponds to roughly one-third of the turnaround radius. The
vertical normalization is arbitrary and has been chosen to match the N-body
results. Taking $(\rho/\sigma^3) \propto \rho^{5/2}/P^{3/2} \propto
(T/\rho^{\gamma-1})^{-5/2\gamma}$ to be a measure of the local `entropy' of the
system, Figure 1a shows that CDM halos have the same radial entropy
stratification as the simple spherical collapse solution. It is possible that
this power-law stratification is a fundamental property which underlies the
similarity of structure of cold dark matter halos.

Density profiles consistent with the power-law phase-space density profile shown
in Figure 1a can be obtained by assuming hydrostatic equilibrium.  The isotropic
Jeans equation admits a family of solutions for the density profile under the
constraint $\rho/\sigma^3 \propto r^{-\alpha}$.  The family is controlled by a
single parameter, which can be expressed in terms of the ratio of the circular
velocity to the 1-D velocity dispersion at $r_p$, the radius where the circular
velocity peaks ($\kappa^{1/2}=V_c(r_p)/\sigma(r_p)$). For $\kappa=\alpha=1.875$,
the family includes a power-law, $\rho \propto r^{-\beta}$, with
$\beta=6-2\alpha=9/4=2.25$ , which corresponds to Bertschinger's spherical
infall solution. As $\kappa$ increases, the density profiles become increasingly
curved, although they still approach the steep power-law divergent behaviour
near the center (see, e.g., second curve from top in Figure 1b). For $\kappa$
greater than some `critical' value, $\kappa_{\rm crit} \simeq 2.678$, the
density profiles become unphysical, vanishing at a finite radius near the
center.

The ``critical'' density profile ($\kappa=\kappa_{\rm crit}$) corresponds to the
maximum value of $\kappa$ consistent with a non-vanishing density profile at the
center and, as shown in Figure 1b, describes the N-body results much better than
a power law. Interestingly, over approximately three decades in radius, {\it the
shape of the NFW density profile is very similar to the `critical' solution}
alluded to above.

The meaning of the critical solution may be understood by considering the
phase-space density distribution corresponding to the different solutions. The
phase-space density distribution function is broadest for the power-law solution
($\kappa=\alpha=1.875$) and gets increasingly narrower as $\kappa$ increases;
the critical solution ($\kappa=\kappa_{\rm crit}$) corresponds then to the most
sharply peaked phase-space density distribution compatible with a monotonic
density profile. In other words, the critical solution may be interpreted as a
``maximally mixed'' configuration where the phase-space density is as uniform as
possible across the system.

This leads to the following interpretation of the origin of the NFW
profile. Gravitational assembly of CDM halos leads to a simple power-law radial
stratification of the phase-space density. If spherical symmetry is imposed, as
in the case treated by Bertschinger (1985), the collapse of each radial mass
shell generates different ``entropies'' (phase-space densities) as they settle
into virial equilibrium, leading to steeply cusped power-law profiles with slope
$\beta=6-2\alpha$. On the other hand, when the assumption of spherical symmetry
is released and the collapse proceeds through many stages of hierarchical
merging, mass shells are continuously ``mixed'' and the profiles tend to the
critical solution: that corresponding to the most uniform entropy distribution
compatible with a monotonic (non-hollowed) density profile and with the
power-law entropy stratification constraint. 

\section{ NFW vs `critical' solution}

\begin{figure}
\plottwo{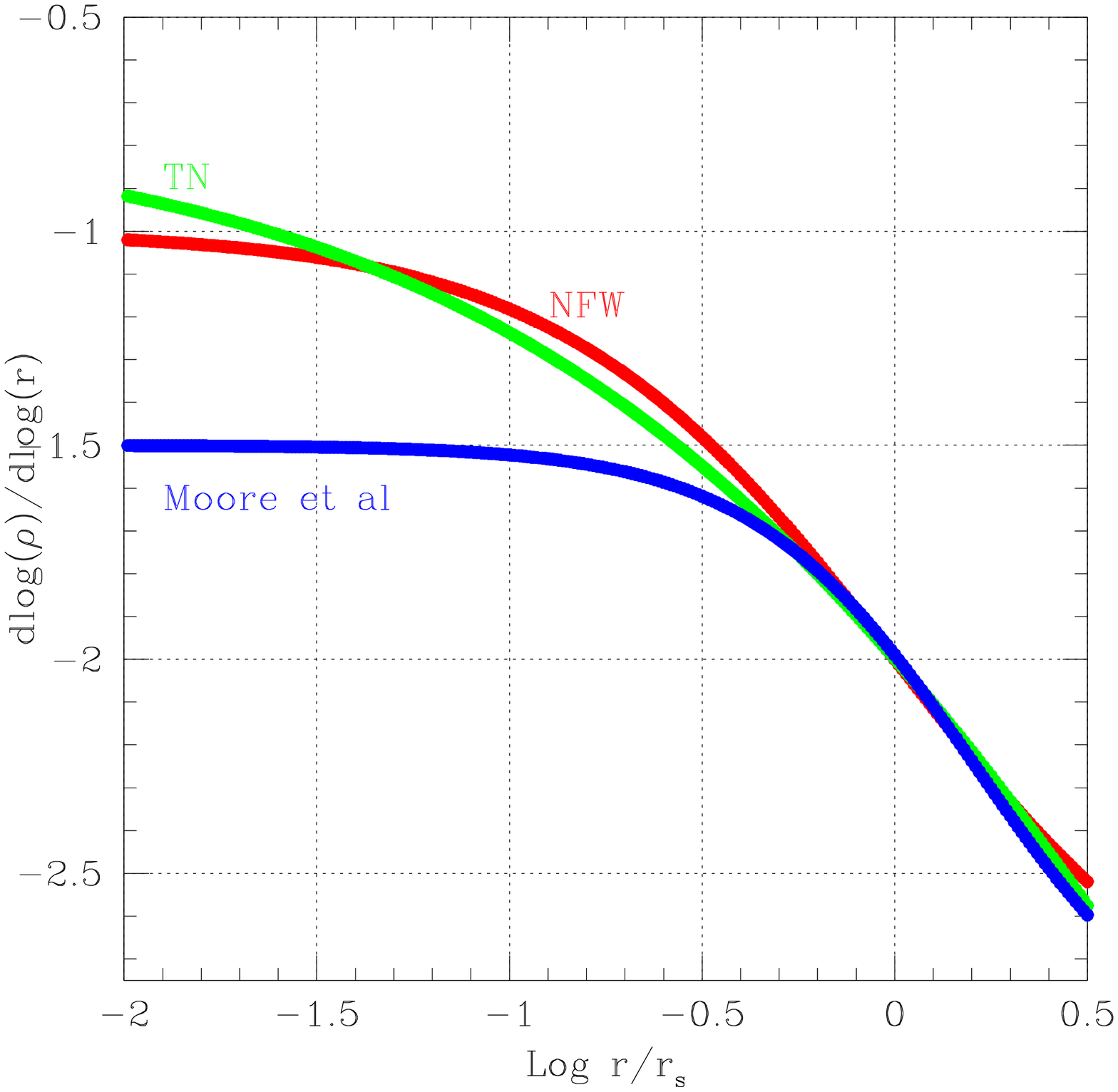}{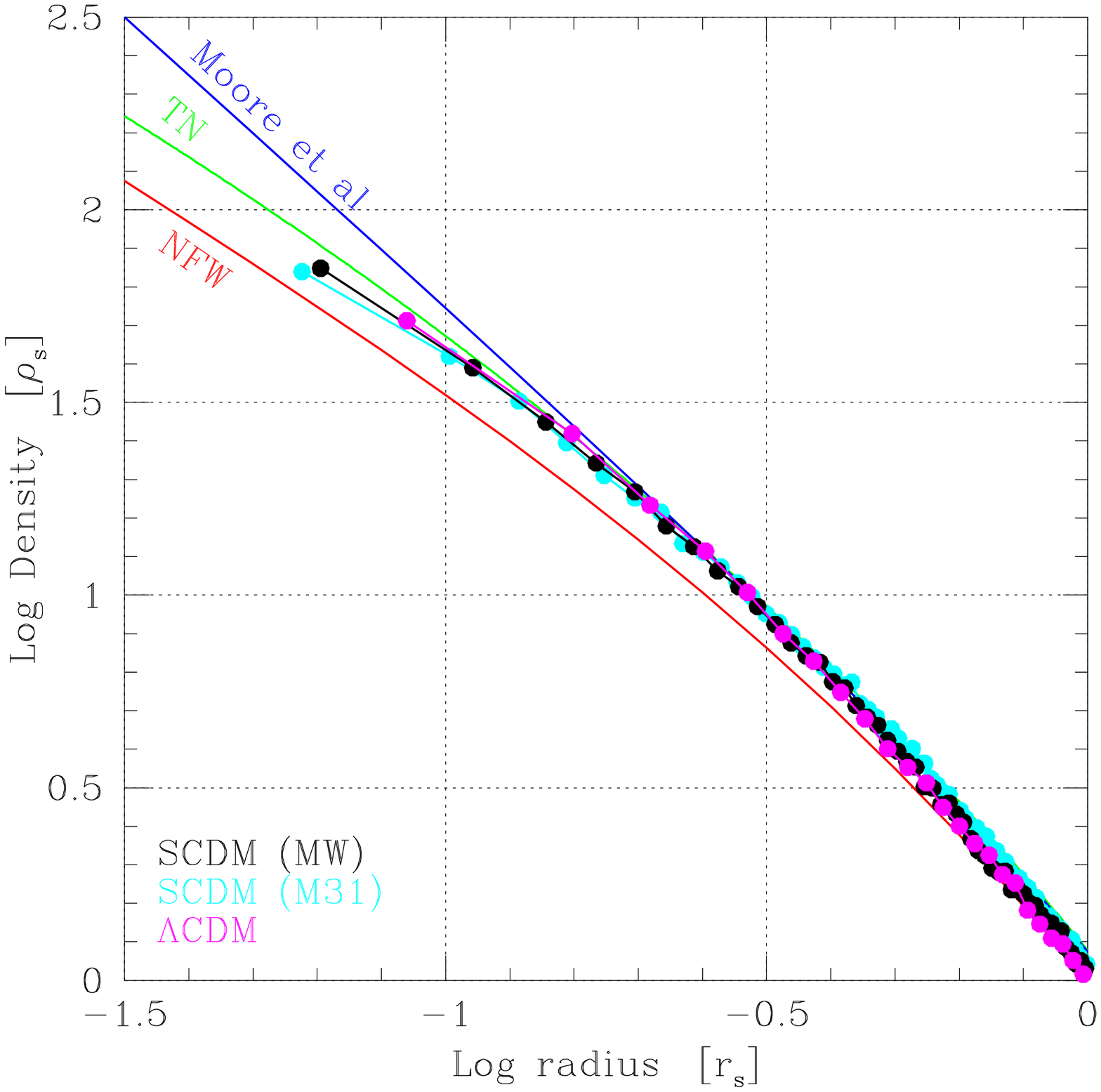}
\caption{
\small  {\it Left (a):} Logarithmic slopes of three different models for the
density profile of dark matter halos as a function of radius.  {\it Right (b):}
Density profiles of three simulated CDM halos (solid circles) compared, from
bottom to top, to the models of NFW, TN, and Moore et al (1998),
respectively. Only radii within the NFW scale radius, $r_s$, are shown in order
to emphasize details of the inner profiles. A significant excess over the NFW
profile is seen for all three systems over the radial range shown, as expected
from the `critical solution' of TN. The Moore et al profile describes the inner
profiles better than NFW in the range $0.15 < r/r_s < 0.5$, but deviates
systematically at smaller radii reliably probed by the simulations.  }
\end{figure}

As discussed above, the `critical' density profile resembles the model proposed
by NFW over a large dynamic range in radius.  Despite this similarity, there are
important differences between the two profiles, one of the most notable being
that the critical solution tends to an asymptotic central slope of $-\beta =
-2\alpha/5 = -0.75$ rather than to $-1$ as in NFW's fitting formula.  A simple
approximation to the radial dependence of the slope of the `critical' density
profile is given by
\begin{equation}
\beta=-{{d\ln \rho}\over{d \ln r}} = {0.75 + 2.625 \, x^{1/2}\over 1+0.5 \, x^{1/2}},
\end{equation}
which is accurate to $3 \% $ for $x = r/r_0 < 4$. Here $r_0=(5/3) r_s$ is the
radius where the logarithmic slope of the density profile equals $-2.25$. This
should be compared with $\beta=(1+3y)/(1+y)$ for NFW, where $y=r/r_s$, or with
the profile proposed by Moore et al (1998), $\beta=-1.5(1+2
x^{-3/2})/(1+x^{-3/2})$. The comparison is shown in Figure 2a. The critical
profile (labeled `TN' in this figure) is overall similar to NFW, although it
becomes shallower inwards more gradually than NFW. For $r\lsim 0.03\, r_s$ the
critical solution is shallower than NFW, and approaches slowly the asymptotic
value of $-0.75$.

These differences are reflected mainly in the inner regions, as illustrated in
Figure 2b. Here the density profile corresponding to the three models are
compared with the simulations{\footnote{Model profiles in Figure 2b have been
fitted only to the outer regions of the simulated halos in order to emphasize
the differences between models. Quantitatively, results vary depending of the
radial range chosen for the fit, but the trends described here remain.}}.  The
three models agree very well at radii exceeding $r_s$, but differ systematically
within $r_s$. Although the critical solution predicts densities larger than NFW
for r just inside $r_s$, the situation eventually reverses for radii smaller
than $\sim 0.01 \, r_s$, outside the range plotted in Figure 2b.

Interestingly, the profiles of simulated halos exhibit similar deviations from
NFW fits over the same radial range: the three curves connecting solid circles,
corresponding to the CDM halos alluded to in \S3, show a clear excess over the
NFW fit. Such excess has been noted by Moore et al (1998), Ghigna et al (2000),
and Fukushige \& Makino (1997, 2001), who interpret it as a signature of very
steep ($\beta \approx -1.5$) central divergent slopes. The modification to the
NFW formula proposed by Moore et al (1998), where the innermost slope is $-1.5$
rather than $-1$, indeed appears to describe the simulation results better than
the NFW formula in the $0.15 < r/r_s < 0.5$ range (see the top curve labeled
`Moore et al' in Figure 2a).

However, and in light of the discussion in \S3, these systematic deviations from
NFW fits do not necessarily signal the onset of a very steep central slope, but
rather may reflect the more gradual radial dependence of the logarithmic slope
of the critical solution (Figure 2a). This interpretation seems to find support
in the simulations: {\it the profiles of the three simulated halos are
significantly shallower than the Moore et al fit near the center}. One may want
to dismiss this feature as an artifact of limited numerical resolution, but
application of our convergence criteria (\S2) to their own data suggest that the
deviations are genuine and point to inner slopes significantly shallower than
$\beta=-1.5$, just as expected from TN's critical solution.

Finally, Figure 3a compares the circular velocity profiles of the same CDM halos
with the predictions of the three models considered in Figure 2. The cumulative
mass profiles of TN's critical solution and NFW are essentially
indistinguishable; the different radial dependence of the density profile in the
two models roughly compensate, and lead to almost identical circular velocity
profiles that are in very good agreement with the results of the
simulations. As expected from the discussion of Figure 2a, the Moore et al model
significantly overestimates the dark mass near the center compared with the
simulations. 

This result is confirmed by a further simulation of a galaxy-sized $\Lambda$CDM
halo, whose density profile is shown in Figure 3b. The profiles correspond to
the halo analyzed by Power et al (2001) for the convergence tests described in
\S2. These tests vary the softening, timestep, and number of particles and show
that extremely high numerical resolution is needed in order to discriminate
between the three models. The highest resolution halo in Figure 3b (which has
more than 3 million particles within $r_{200}$) has a circular velocity profile
which agrees well with NFW (and TN's critical solution) but is much shallower
near the center than expected from Moore et al's fitting formula. 

Can these results be used to place meaningful constraints on the asymptotic
inner slope?  At $r_{\rm min}\sim 1\, h^{-1}$ kpc, the smallest radius resolved
in the highest-resolution run shown in Figure 3b, both the local and cumulative
density profiles are robustly determined: $\rho(r_{\rm min})/\rho_{\rm
crit}=9.4\times 10^5$, and ${\bar \rho}(r_{\rm min})/\rho_{\rm crit}\approx
1.6\times 10^6$. These values can be combined with the requirement of mass
conservation to place an upper limit to the inner asymptotic slope of the
density profile, $\beta < 3 (1-\rho(r_{\rm min})/{\bar \rho(r_{\rm
min})})=1.2$. In other words, there is not enough mass within $r_{\rm min}$ to
support a power-law density profile with slope steeper than $\beta=1.2$. We note
that this conclusion depends sensitively on our ability to resolve the innermost
$1\, h^{-1}$ kpc. If $r_{\rm min}$ were just two or three times larger the same
exercise would not be able to rule out slopes as steep as $\beta=1.5$.

\section{Summary and Discussion}

\begin{figure}
\plottwo{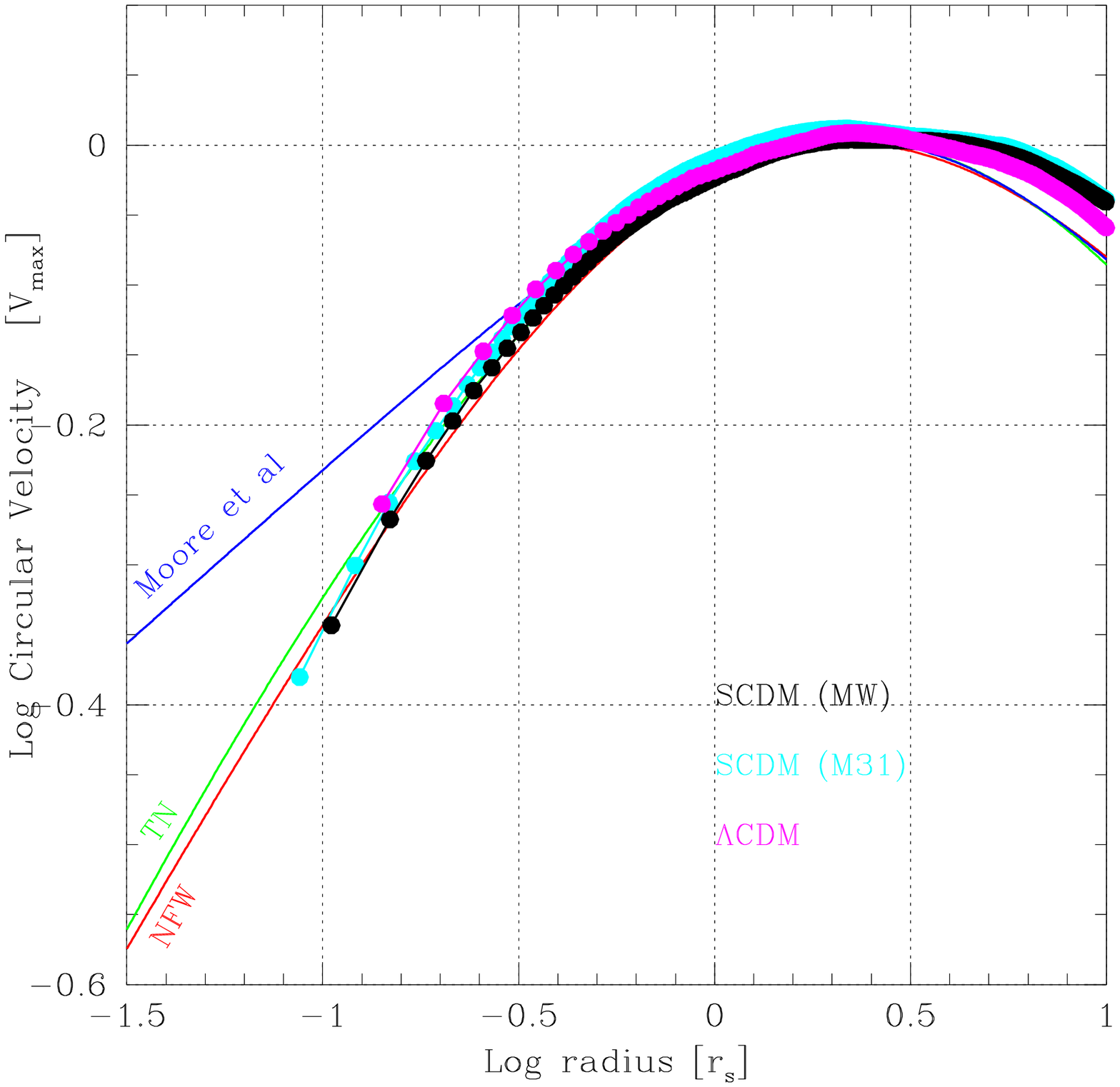}{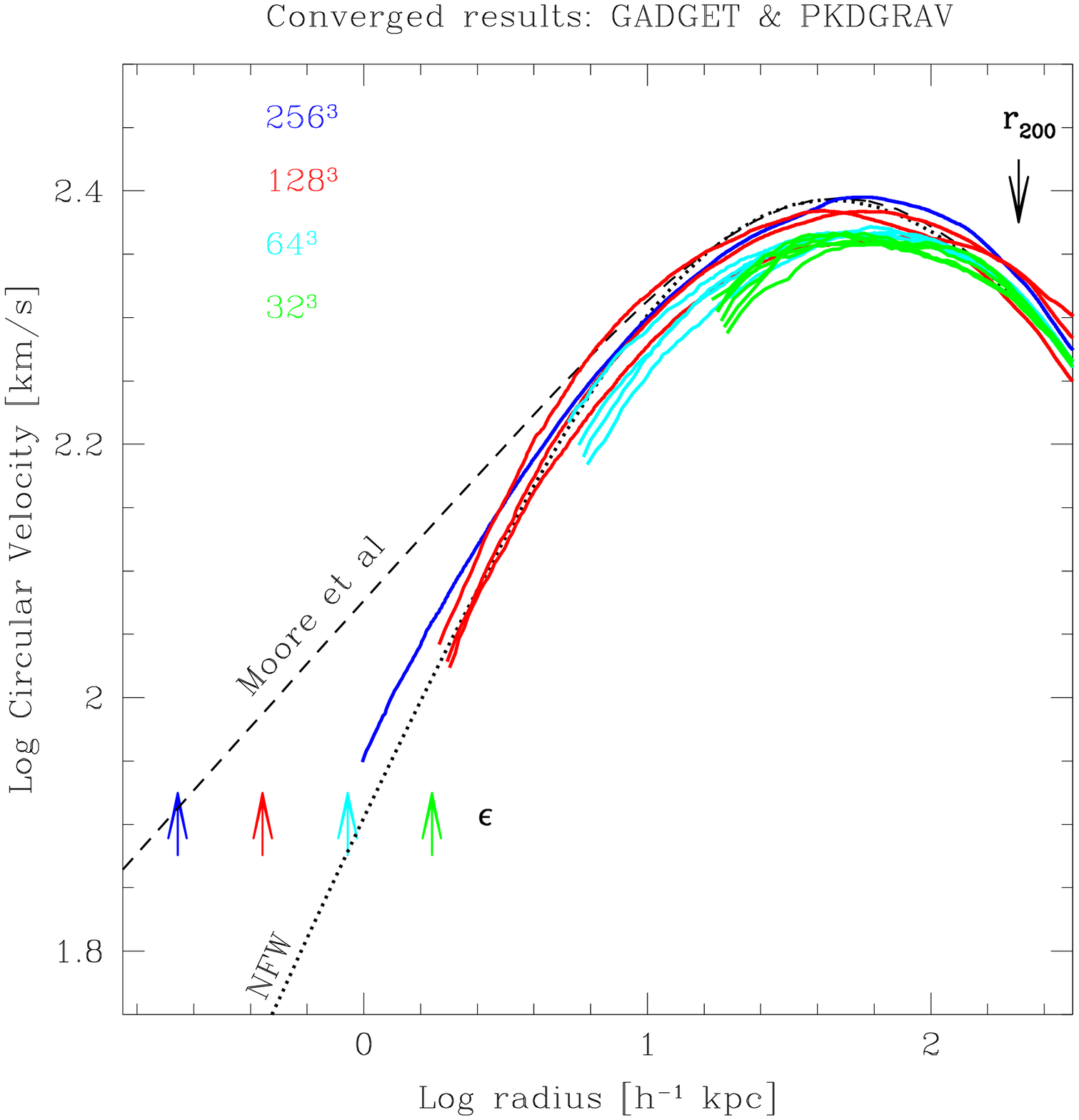}
\caption{\small
{\it Left (a):} As in Figure 2b, but for the circular velocity profiles. Because
TN's `critical solution' is shallower than NFW near the center and steeper
further out, there is little difference between the circular velocity profiles
of these two models, and both seem to reproduce well the results of the
simulations down to the innermost resolved radius. The Moore et al profile
overestimates significantly the amount of dark mass near the center.  {\it Right
(b):} Circular velocity profiles of a $\Lambda$CDM halo taken from the
convergence study of Power et al (2001). Profiles are plotted for a large number
of `converged' runs with different number of particles and are shown for radii
where the convergence criteria summarized in \S2 are satisfied. The value of the
gravitational softening is indicated by the small vertical arrows. The
convergent circular velocity profile that emerges from all runs is roughly
independent of the number of particles and resembles closely the model proposed
by NFW and the `critical' solution of TN.  Steeply-cusped density profiles such
as that proposed by Moore et al (1998) are disfavored by these data.}
\label{figs:vprof}
\end{figure}

In summary, a number of interesting features emerge from the analysis of four
very high-resolution (million-particle class) simulations of galaxy-sized CDM
halos.
\begin {itemize}
\item
The spherically-averaged, phase-space density profiles of CDM halos are very
well approximated by a power law, $\rho/\sigma^3 \propto r^{-1.875}$. The slope
of this power law is consistent with the self-similar solution for spherical
secondary infall derived by Bertschinger (1985).
\item
The circular velocity profiles of simulated CDM halos agree well with the model
proposed by NFW down to the smallest resolved scales.
\item
On the other hand, the shape of the density profiles of simulated CDM halos
differs modestly, but significantly, from the model proposed by NFW. As a
result, NFW fits tend to underestimate the density just inside the scale radius
$r_s$, as pointed out by Moore et al (1998), Ghigna et al (2000), and Fukushige
\& Makino (1997, 2001).
\item
There is no obvious evidence for convergence to a steep asymptotic slope over
the radial range probed by the simulations. Density profiles get increasingly
shallow near the center, with innermost slopes significantly shallower than
$-1.5$.
\item
A maximally-mixed (`critical') model where the phase-space density profile is
assumed to be a power law of radius accounts for these features and predicts
that the slope of the density profile should gradually approach $-0.75$ near the
center.
\end{itemize}

The power-law behaviour of the phase-space density thus offers a natural way to
gain insight into the structure of dark matter halos at radii where simulations
become increasingly difficult and expensive, but also where observational
constraints are strongest. Provided that the velocity dispersion tensor remains
nearly isotropic, the ``critical'' solution provides a clear prediction as to
the behaviour of the logarithmic slope of the density profile: it should become
progressively shallower towards the center, converging to a value of
$-\beta=-0.75$.

This is interesting, because slopes shallower than $-1.5$ appear to be
consistent with the recent reanalysis of the rotation curve dataset by van den
Bosch \etal (2000) and by van den Bosch \& Swaters (2001). It is important to
mention, however, that a shallow central slope does not guarantee consistency
with observations, which constrain the detailed radial dependence of the density
profile slope as well. Our results do, however, offer a clear prediction to
extrapolate the mass profiles to regions that are very difficult to probe
numerically.  Is the extrapolation of the power-law behaviour to very small
radii warranted?  This question ultimately will have to be answered by direct
numerical simulation, although there is no obvious a priori reason why a power
law scaling that is valid for over two decades in radii should break down nearer
the center.

We note that these conclusions are at odds with the proposal of Moore et al
(1998) and Ghigna \etal (2000) that the innermost slopes of CDM halos converge
to a value not shallower than about $-1.5$. However, it should be emphasized
that their conclusion was based on the simulation of a {\it single} halo
simulated in a standard CDM universe (SCDM) and on a significantly different
mass regime (galaxy clusters) than probed here. Thus the possibility remains
that this particular system may not be representative of the general population
or that the density profiles of clusters are steeper than those of galaxy-sized
halos.  Confirming which of these possibilities holds will require a
statistically significant sample of halos simulated with resolution comparable
to the systems used here. Finally, it is also possible that the cluster
simulated by Moore \etal (1998) and Ghigna \etal (2000) differs from the
galaxy-sized halos we present here in other, more subtle ways. For example, it
may be significantly more triaxial than the systems analyzed here, or perhaps
its velocity dispersion tensor is very anisotropic, in conflict with the
assumptions of this work. Again, a detailed reanalysis of the discrepant system,
extended to a statistically meaningful sample, appears necessary in order to
explain conclusively this discrepancy.

The good agreement between the `critical' solution and the mass profiles of
simulated halos suggests that the structure of CDM halos is determined by a
radial stratification of phase-space density similar to that established through
collapse onto a point mass perturbation in an unperturbed expanding universe,
and by the uniformization of phase-space density that occurs presumably as a
result of the many merger and satellite accretion events that characterize the
assembly of a CDM halo. This identification leaves a couple of important
questions unanswered: (i) why should the phase-space density be a power law of
radius?, and (ii) why is the exponent the same as in Bertschinger's self-similar
solution?. Although we have no clear answer to these questions at this point,
our results suggest that explaining the origin of the structural similarity of
CDM halos pointed out by Navarro, Frenk \& White may entail unraveling why the
radial stratification of phase-space density in CDM halos is the same power-law
generated by the simple spherical collapse model. Discovering a mechanism that
achieves this may provide a simple explanation for the universal structure of
cold dark matter halos.

\acknowledgements
I am grateful to my collaborators: Carlos Frenk, Adrian Jenkins, Chris Power,
Tom Quinn, Volker Springel, Joachim Stadel, James Taylor, and Simon White, for
allowing me to report our results here, and also to J.Makino, T.Fukushige and
P.Hut for organizing a very enjoyable meeting. Cool talks, fresh sushi, cold
Sapporo, great company: all seemed a perfect antidote for a hot July in Tokyo
and Kyoto. Thanks as well to B.Moore, S.Ghigna, G.Lake, and F.Governato for
sharing with me data from their `Local Group' simulations. This work has been
supported by various grants from NSERC and CFI.


\end{document}